\documentclass[showpacs,10pt,twocolumn,prb]{revtex4}

\usepackage{graphicx}
\usepackage{epsfig}

\begin{document}

\title{Weak Ferromagnetism in Fe$_{1-x}$Co$_{x}$Sb$_{2}$}
\author{Rongwei Hu$^{1,2}$, R. P. Hermann$^{3\ast }$, F. Grandjean$^{3}$, Y.
Lee$^{4}$, J. B. Warren$^{5}$, V. F. Mitrovi{\'{c}}$^{2}$ and C. Petrovic$%
^{1}$}
\affiliation{$^{1}$Condensed Matter Physics, Brookhaven National Laboratory, Upton New
York 11973-5000 USA\\
$^{2}$Physics Department, Brown University, Providence RI 02912\\
$^{3}$Department of Physics B5, Universit\'{e} de Li\`{e}ge, Belgium\\
$^{4}$Department of Earth System Sciences, Yonsei University, Seoul 120749,
Korea\\
$^{5}$Instrumentation Division, Brookhaven National Laboratory, Upton New
York 11973-5000 USA}
\date{\today}

\begin{abstract}
Weak ferromagnetism in Fe$_{1-x}$Co$_{x}$Sb$_{2}$ is studied by
magnetization and M\"{o}ssbauer measurements. A small spontaneous magnetic
moment of the order of $\sim 10^{-3}\mu _{B}$ appears along the $\widehat{b}$%
-axis for $0.2\leq x\leq 0.4$. Based on the structural analysis, we argue
against extrinsic sources of weak ferromagnetism. We discuss our results in
the framework of the nearly magnetic electronic structure of the parent
compound FeSb$_{2}$.
\end{abstract}

\maketitle

\section{Introduction}

FeSi and FeSb$_{2}$ are semiconductors that show crossover from a
nonmagnetic semiconducting ground state with a narrow gap to a thermally
induced paramagnetic metal with enhanced susceptibility. \cite{Hulliger}$%
^{,} $\cite{Petrovic1} The magnetic properties of FeSi have instigated
considerable theoretical interest, starting with the narrow-band model of
Jaccarino. \cite{Jaccarino} Further models include a nearly ferromagnetic
semiconductor model of Takahashi and Moriya\cite{Takahashi} in which the
state was sustained by thermally induced spin fluctuations found in neutron
scattering experiments. \cite{Shirane}$^{,}$\cite{Tajima} Moreover, the
nearly ferromagnetic semiconductor picture was supported by \textit{LDA+U}
band structure calculations by Mattheiss and Hamann\cite{Mattheiss} and
Anisimov \textit{et al}. \cite{Anisimov1} At the same time, Aeppli and Fisk%
\cite{FiskAeppli} pointed out that the magnetic properties of FeSi are
analogous to the physics of Kondo insulators, albeit with a reduced on-site
Coulomb repulsion $U$. The basis of their argument was a model, ruled out by
Jaccarino in his original work, of the narrow gap and high density of
states. Experiments of Mandrus \textit{et al}.\cite{Mandrus} and Park 
\textit{et al}.\cite{Park} confirmed the validity of the model of Aeppli and
Fisk.

A search for new model systems, where the applicability of the Kondo
insulator framework to \textit{3d} transition metals can be investigated,
led to the synthesis of large single crystals of FeSb$_{2}$. Furthermore, a
crossover was discovered similar to the one in FeSi, for the magnetic and
electrical transport properties. \cite{Petrovic1}$^{,}$\cite{Petrovic2}
Subsequent alloying studies have shown heavy fermion metallic state induced
in FeSb$_{2-x}$Sn$_{x}$, just as in FeSi$_{1-x}$Al$_{x}$. \cite{Danes}$^{,}$%
\cite{DiTusa} In both materials the optical conductivity revealed
unconventional charge gap formation. That is, a complete recovery of
spectral weight in FeSi and FeSb$_{2}$ occurs over an energy range of few
eV, suggesting contributions of larger energy scales. \cite{Schlesinger,Leo}
This is in sharp contrast to metal-insulator transitions in band insulators
where thermal excitations of charge carriers through the gap redistribute
just above the gap.

One of the key predictions of the \textit{LDA+U} approach was the close
proximity of FeSi to a ferromagnetic state. \cite{Anisimov2} In analogy to
FeSi, recent ab-initio calculations predicted the nearly ferromagnetic
nature of the FeSb$_{2}$ ground state. \cite{Lukoyanov} In FeSi the
ferromagnetic state has been induced by lattice expansion in FeSi$_{1-x}$Ge$%
_{x}$ \cite{Sunmog} or by carrier insertion in Fe$_{1-x}$Co$_{x}$Si. \cite%
{Dave} In contrast, FeSb$_{2}$ has not yet been tuned to a ferromagnetic
state by any external parameters. In this work, we demonstrate the presence
of the weak ferromagnetism (WFM) in Fe$_{1-x}$Co$_{x}$Sb$_{2}$ ($0.2\leq
x\leq 0.45$). The origins of the WFM are discussed. Extensive structural
analysis shows no evidence of extrinsic impurity induced WFM. We argue that
instead the WFM is a consequence of the nearly ferromagnetic electronic
structure of the parent compound FeSb$_{2}$.

\section{Experiment}

The Fe$_{1-x}$Co$_{x}$Sb$_{2}$ single crystals were grown from excess Sb
flux. \cite{Petrovic1} Powder X-ray diffraction (XRD) patterns of the ground
samples were taken with Cu K$_{\alpha }$ radiation ($\lambda =1.5418$ \AA )
using a Rigaku Miniflex X-ray diffractometer. The lattice parameters were
obtained using Rietica software. \cite{Hunter} High resolution XRD patterns
were taken at the beamline X7A of the National Synchrotron Light Source at
the Brookhaven National Laboratory using monochromatic synchrotron X-ray and
gas-proportional position-sensitive detector. Rietveld refinements were
performed using GSAS.\cite{GSAS} A JEOL JSM-6500 SEM microprobe with
resolution of 1.5 nm was used for verifying the Co concentrations and
investigating the microstructure. Single crystals were oriented using a Laue
Camera. Magnetization measurements were performed in a Quantum Design MPMS
XL 5 instrument. The iron-57 M\"{o}ssbauer spectra, at temperatures ranging
from 2.8 to 295 K, were measured on a constant acceleration spectrometer
that utilized a rhodium matrix cobalt-57 source. The instrument was
calibrated at 295 K with $\alpha $-iron powder. The isomer shifts reported
herein are relative to $\alpha $-iron at 295 K. The thickness of the
absorber was 23 and 72 mg/cm$^{2}$ for FeSb$_{2}$ and Fe$_{0.75}$Co$_{0.25}$%
Sb$_{2}$, respectively. The sample temperature in the Janis SV-300 cryostat
was controlled with a LakeShore 330 temperature controller and a silicon
diode mounted on the copper sample holder. The accuracy of the sample
temperature is better than $\pm 1$\%.

The powder X-ray patterns show that the Fe$_{1-x}$Co$_{x}$Sb$_{2}$ ($0.2\leq
x\leq 0.45$) samples crystallize in the \textit{Pnnm} structure without any
additional crystalline peaks introduced by Co alloying. The effect of Co
substitution on the Fe site is to expand the unit cell volume as compared to
FeSb$_{2}$. This expansion is anisotropic and results from a contraction in
the basal \textit{a-b} plane and an expansion along the \textit{c}-axis upon
substitution of Fe by Co. \cite{Rongwei}

\section{Magnetic properties}

\begin{figure}[thb]
\centerline{\includegraphics[scale=0.65]{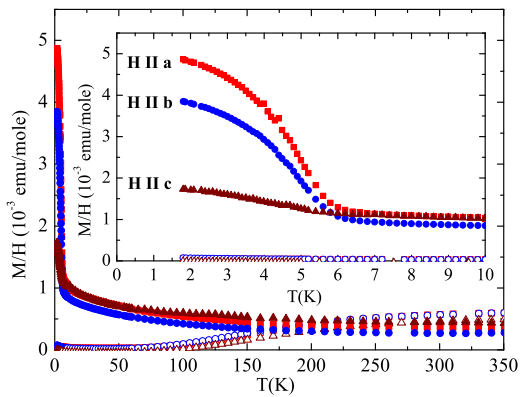}} 
\vspace*{-0.2cm}
\caption{Magnetic susceptibility M/H of FeSb$_{2}$ (open symbols)\ and Fe$%
_{0.75}$Co$_{0.25}$Sb$_{2}$ (filled symbols) for a 1kOe field applied along
all three principal crystalline axes.}
\label{Fig1}
\end{figure}

At low temperature, the parent compound FeSb$_{2}$ is a narrow gap
semiconductor with a rather small and temperature independent magnetic
susceptibility. \cite{Petrovic1} Similar to FeSi, above 100 K there is a
temperature induced paramagnetic susceptibility and an enhanced electronic
conduction. The magnetic susceptibility can be described by both a thermally
induced Pauli susceptibility and a low to high spin transition. \cite%
{Petrovic1}$^{,}$\cite{Petrovic2}$^{,}$\cite{Grandjean} In the temperature ($%
T$) range from 1.7 to 150 K the Fe$_{0.75}$Co$_{0.25}$Sb$_{2}$ magnetic
susceptibility is larger than that of FeSb$_{2}$. For $T$ above 6 K, it
shows little anisotropy with the magnetic field applied along the different
crystallographic axes. As shown in Fig. 1, the temperature dependence of the
susceptibility indicates Pauli paramagnetism at high temperature. A clear
ferromagnetic transition at $T_{C}=6$ K for a field of 1 kOe applied along
any of the three crystallographic axes is illustrated in the inset to Fig.
1. These observations are in agreement with ferromagnetic long range order
of the small magnetic moments below $T=5$ K.\cite{Rongwei} The ferromagnetic
nature of the transition is supported by the hysteresis loop measured at $%
T=1.8$ K and displayed in Fig. 2. For field strength varying between -6 and
6 kOe applied along the $\widehat{\mathit{b}}$ - axis, hysteresis loops are
observed for $0.20\leq x\leq 0.45$. The width of the hysteresis loop grows
initially with increasing $x$ from $x=0.20$, peaks at $x=0.25$, and becomes
progressively smaller upon further Co substitution. Hysteresis loops are
absent for field applied along the $\widehat{\mathit{c}}$ - axis and are
observed only for $x=0.25$ for field applied along the $\widehat{\mathit{a}}$
- axis. By extrapolating the magnetization of Fe$_{0.75}$Co$_{0.25}$Sb$_{2}$
to H=0, a lower estimate of the saturation magnetization along the b-axis of
M$_{LL}$ = $0.0005$ $\mu _{B}$/F.U.) or ($5.10^{-4}$ $\mu _{B}$/Fe), where
F.U. refers to the FeSb$_{2}$ formula unit, is obtained.

\begin{figure}[thb]
\centerline{\includegraphics[scale=0.65]{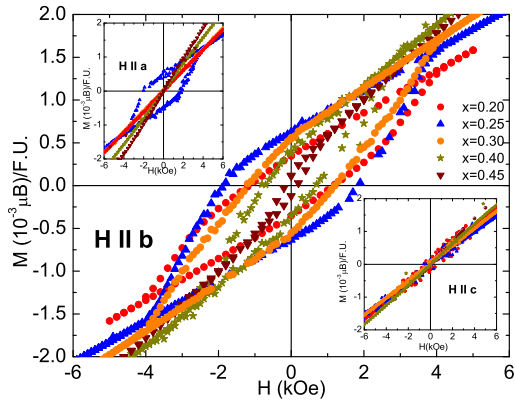}} 
\vspace*{-0.2cm}
\caption{Hysteresis loops for Fe$_{1-x}$Co$_{x}$Sb$_{2}$ in the
ferromagnetic state $(x=0.2-0.45)$ at T =1.8 K. Magnetization does not
saturate, it continues to increase with applied magnetic field, similar to
bulk itinerant ferromagnets with \textit{3d} ions.}
\label{Fig2}
\end{figure}

\begin{figure}[thb]
\centerline{\includegraphics[scale=0.65]{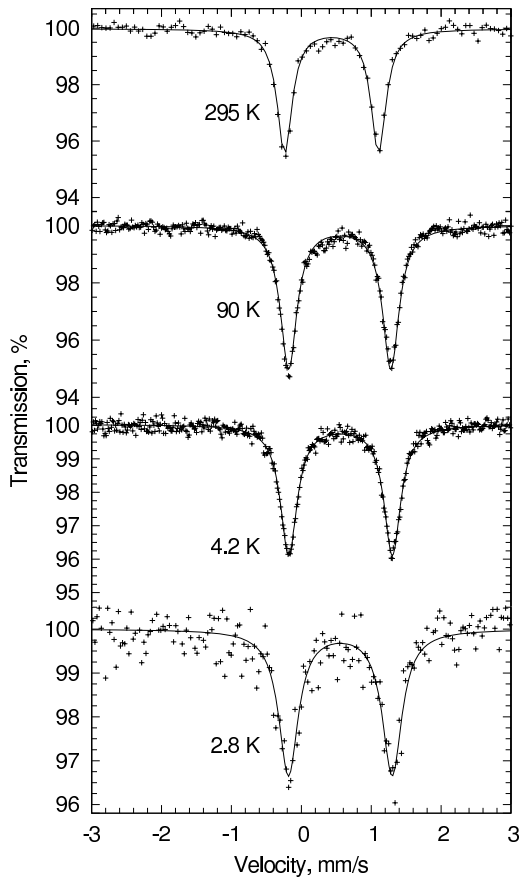}} 
\vspace*{-0.2cm}
\caption{The $^{57}$Fe M\"{o}ssbauer spectra of Fe$_{0.75}$Co$_{0.25}$Sb$_{2}
$ at the indicated temperatures. The solid line is a fit to a doublet, using
the parameters indicated in Table 1.}
\label{Fig3}
\end{figure}

\begin{table*}[tph]
\caption{The hyperfine parameters for FeSb$_{2}$ and Fe$_{0.75}$Co$_{0.25}$Sb%
$_{2}\,^{a}$; $^{\mathbf{a}}$Relative to $\protect\alpha $-iron at 295K and $%
^{\mathbf{b}}$The isomer shift reference in Ref. 24 is sodium nitroprusside,
which has a -0.26 mm/s isomer shift relative to $\protect\alpha $-iron at
room temperature.}
\label{tbl1}%
\begin{tabular}{|l|l|l|l|l|l|l|}
\hline
\ \ \ \  & \multicolumn{3}{|l}{Fe$_{0.75}$Co$_{0.25}$Sb$_{2}$} & 
\multicolumn{3}{|l|}{FeSb$_{2}$} \\ \hline
T(K) & $\delta ^{a}$,mm/s & $\Delta $E$_{Q}$,mm/s & $\Gamma $,mm/s & $\delta
^{a}$,mm/s & $\Delta $E$_{Q}$,mm/s & $\Gamma $,mm/s \\ \hline
296$^{b}$ & - & - & - & 0.450(6) & 1.286(6) & - \\ \hline
295 & 0.433(2) & 1.343(2) & 0.264(5) & 0.449(1) & 1.275(2) & 0.262(3) \\ 
\hline
240 & 0.483(2) & 1.394(2) & 0.265(2) & - & - & - \\ \hline
190 & 0.509(2) & 1.422(2) & 0.267(2) & - & - & - \\ \hline
140 & 0.538(2) & 1.449(2) & 0.273(2) & - & - & - \\ \hline
90 & 0.455(2) & 1.364(2) & 0.284(2) & - & - & - \\ \hline
50 & 0.569(2) & 1.474(3) & 0.290(4) & - & - & - \\ \hline
6.4$^{b}$ & - & - & - & 0.572(6) & 1.575(6) & - \\ \hline
4.2 & 0.560(2) & 1.483(2) & 0.291(2) & 0.572(1) & 1.573(3) & 0.270(4) \\ 
\hline
2.8 & 0.558(2) & 1.483(3) & 0.338(4) & - & - & - \\ \hline
\end{tabular}%
\end{table*}

The M\"{o}ssbauer spectra of FeSb$_{2}$ single crystals exhibit a doublet at
T=295 and 4.2 K. No impurity, and in particular no impurity with a large
hyperfine field, is observed in the M\"{o}ssbauer spectra. Furthermore, the M%
\"{o}ssbauer spectral parameters for FeSb$_{2}$ obtained herein are in
excellent agreement with the previously reported parameters (see Table I). 
\cite{Raphael} For Fe$_{0.75}$Co$_{0.25}$Sb$_{2}$, the M\"{o}ssbauer
spectra, shown in Fig. 3, exhibit a doublet for temperatures ranging from $%
T=295$ K to 2.8 K. Again no impurity contribution is observed. Its spectral
parameters, obtained at $T=295$ K and 4 K, are close to those observed in
the FeSb$_{2}$. The isomer shift observed in Fe$_{0.75}$Co$_{0.25}$Sb$_{2}$
is ca. 0.01 mm/s smaller than in FeSb$_{2}$. This indicates a somewhat
larger s-electron density at the $^{57}$Fe nucleus.

\begin{figure}[thb]
\centerline{\includegraphics[scale=0.65]{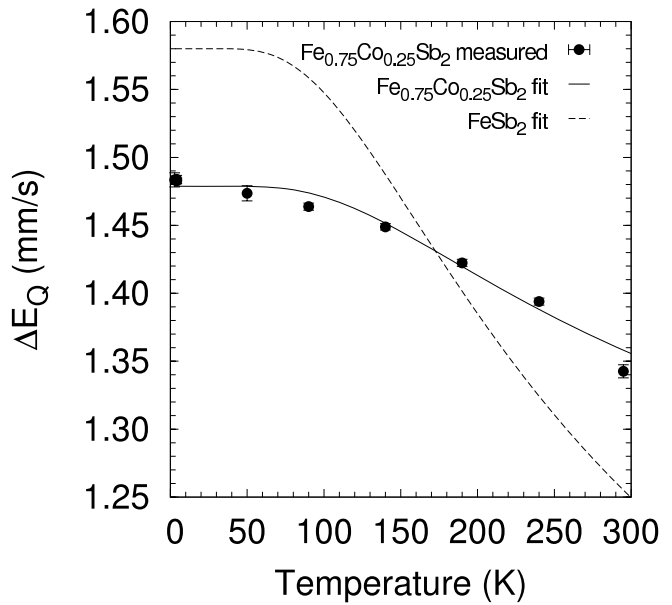}} 
\vspace*{-0.2cm}
\caption{Fit of the Fe$_{0.75}$Co$_{0.25} $Sb$_{2}$ quadrupole splitting as
a function of temperature with the delocalization model.}
\label{Fig4}
\end{figure}

The variation of the quadrupole splitting from 295 to 4.2 K is larger in FeSb%
$_{2}$ than in Fe$_{0.75}$Co$_{0.25}$Sb$_{2}$. This strong temperature
dependence of the quadrupole splitting in FeSb$_{2}$ is consistent with a
scenario of electron delocalization appearing with increasing temperature,
with a gap $\Delta $E of 380 K. \cite{Grandjean}$^{,}$\cite{Goodenough} As
illustrated in Fig. 4, a fit of the Fe$_{0.75}$Co$_{0.25}$Sb$_{2}$
quadrupole splitting as a function of temperature with the delocalization
model described in Ref. 24 yields a somewhat larger gap energy $%
E_{g}=(480\pm 50)$ K than that observed in FeSb$_{2}$. The difference
between the hyperfine parameters in FeSb$_{2}$ and Fe$_{0.75}$Co$_{0.25}$Sb$%
_{2}$ indicates that there is indeed a modification of the FeSb$_{2}$
structure, and that no phase segregation is present. The Mossbauer spectra
show that the investigated phase is (Fe,Co)Sb$_{2}$ and not FeSb$_{2}$+CoSb$%
_{2}$ since the hyperfine parameters are significantly different.
Furthermore no iron-bearing impurity is observed in Fe$_{0.75}$Co$_{0.25}$Sb$%
_{2}$.

Apparently, the T = 2.8 K spectrum of Fe$_{0.75}$Co$_{0.25}$Sb$_{2}$ is a
doublet, which is somewhat surprising. Our interpretation is that either the
iron experiences no magnetic hyperfine field or that the hyperfine field is
below the detection limit. If the small broadening of ca. 0.047(6) mm/s of
the 2.8 K spectrum, when compared to the 4.2 K spectrum, was associated to a
magnetic hyperfine field, it would correspond to a 1.5+/-0.2 kOe hyperfine
field. With a linewidth constrained to 0.29 mm/s, a fit of this spectrum,
with both a quadrupole interaction and a hyperfine field yields a field of
2.8+/-1.2 kOe. Taking the usual proportionality of ca. 150 kOe/$\mu _{B}$,
these values can be used to estimate an upper limit of about M$_{UL}$ = 0.01 
$\mu _{B}$. for the magnetic moment on Fe.

\begin{figure}[thb]
\centerline{\includegraphics[scale=0.9]{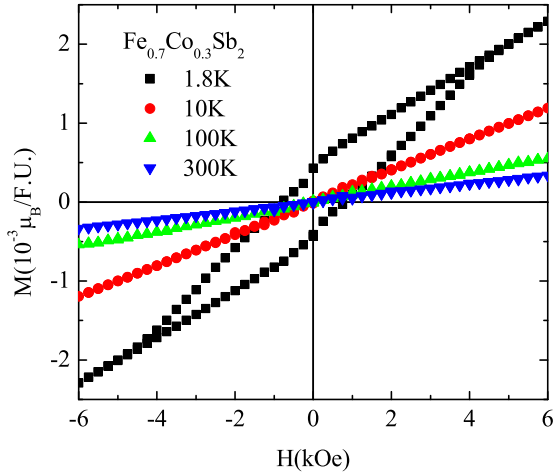}} 
\vspace*{-0.2cm}
\caption{Magnetic hysteresis for the Fe$_{0.7}$Co$_{0.3}$Sb$_{2}$ sample. It
is important to note the absence of hysteresis loops above the ferromagnetic
transition of FeSb$_{2}$. Data were taken on a crystal from independently
grown batch.}
\label{Fig5}
\end{figure}

\section{Intrinsic vs. Extrinsic Magnetism}

Given the small value of the saturated moment, it is possible that WFM
originates from extrinsic sources, such as artifacts of the measurement
process or the presence of a small amount of ferromagnetic impurity, e.g.
elemental Fe. The former can be excluded based on the lack of sample
dependence, both in magnetization and in heat capacity data. \cite{Rongwei}
Below we discuss the possibility of undetected second phases as extrinsic
sources of the WFM.

\begin{figure}[thb]
\centerline{\includegraphics[scale=0.7]{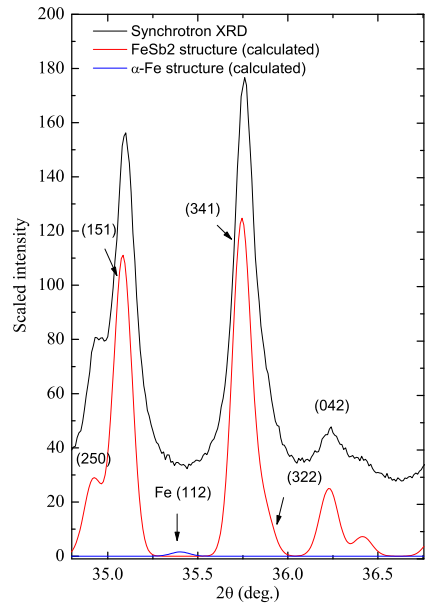}} 
\vspace*{-0.2cm}
\caption{Observed (black) and calculated (red) synchrotron powder X - ray
diffraction patterns of Fe$_{0.75}$Co$_{0.25}$Sb$_{2}$. Calculated pattern
includes 0.3 \% of superimposed $\protect\alpha $ - Fe impurity. If present,
impurity would have caused detectable deviation of the observed pattern
since there is no peak overlap.}
\label{Fig6}
\end{figure}

No hysteresis loops are observed for temperatures above $T_{C}=(6-7)K$ for $%
x=(0.2-0.4)$ (example shown in Fig. 5). No known Fe-Sb, Co-Sb, Fe-Co, Fe-O,
or Co-O phases show a ferromagnetic transition in this temperature range.
FeCo alloys have large hyperfine fields (200-400 kOe) that would have been
detected by M\"{o}ssbauer measurement. We can calculate the X-ray patterns
expected in the presence of bulk crystalline Fe impurities by superimposing
the strongest peak of 0.3\% elemental Fe to the measured patterns. No
overlap between the calculated and measured Fe$_{0.75}$Co$_{0.25}$Sb$_{2}$
X-ray patterns was observed (Fig. 6). Any other unknown Fe-O, Fe-Co-O, Co-O,
Fe-Co, Fe-Sb-Co, \textit{etc.} phase with the same atomic ratio in the
mixture would have been detected and refined by synchrotron powder X-ray
diffraction because its contribution to the scattering mixture would be
higher than that of Fe. Though $M_{LL}$ observed in magnetic hysteresis
loops could be caused by Fe impurities of the order of the synchrotron
powder X - ray diffraction detection limit, absence of hysteresis loops
above 6 K strongly argues against such scenario.

\begin{figure}[thb]
\centerline{\includegraphics[scale=0.4]{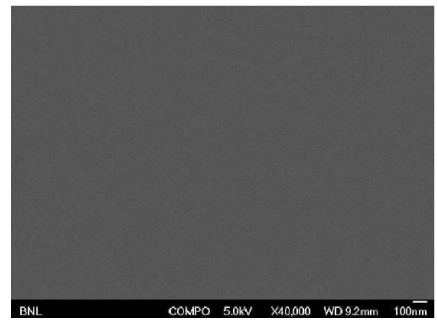}} 
\vspace*{-0.2cm}
\caption{Typical Scanning Electron Microscope (SEM) image of Fe$_{0.75}$Co$%
_{0.25}$Sb$_{2}$. SEM images on randomly chosen surface did not detect
secondary phases or randomly distributed nanoparticles.}
\label{Fig7}
\end{figure}

Another possibility is that magnetism in Fe$_{1-x}$Co$_{x}$Sb$_{2}$ is
caused by magnetic nanoparticles. M\"{o}ssbauer measurement\ shows no
evidence of iron bearing nanoparticles (either FeCo or Fe-oxide). Such
nanoparticles would have a paramagnetic spectrum with different isomer shift
and quadrupole splitting at room temperature, which would be detected with a
0.3\% limit. Below the nanoparticle blocking temperature the field would be
large, typically 500 kOe for typical oxides. Solid evidence against
nanoparticles or bulk extrinsic phases comes from energy dispersive scanning
electron microscope (SEM) measurements. Among the samples grown from several
different batches for $x=0.25$, the uncertainty in Co concentration is $%
x=0.04$. SEM data taken with resolution down to 1.5 nm exclude the presence
of either bulk secondary phases or embedded nanoparticles. This is because
high resolution SEM images of several randomly chosen polished crystals and
crystalline surfaces show no trace of nanosize inclusions, clusters or
inhomogeneities (example shown in Fig. 7). The images were taken in the
\textquotedblleft composition\textquotedblright\ mode with a solid state
detector consisting of paired PN junctions. This type of detector is very
sensitive to back-scattered electrons which in turn are sensitive to local
variations in atomic number. If nano-crystallites of Fe or other elements
were present, they would have been visible as bright dots in the high
magnification image.

Taken in conjunction, our results argue against extrinsic sources of WFM in
Fe$_{1-x}$Co$_{x}$Sb$_{2}$. Recent muon spin relaxation measurements
indicate that the WFM state is spread throughout the full sample volume for
Fe$_{0.7}$Co$_{0.3}$Sb$_{2}$, further supporting our results.\cite{Uemura}

\section{Discussion and Conclusions}

Examples of intrinsic WFM states in narrow band materials are abundant in
nature. \cite{DeLong} Besides numerous oxide compounds, many intermetallic
systems also exhibit intrinsic weak ferromagnetism, such as YbRhSb, \cite%
{YbRhSb} MnS, \cite{MnS} and Yb$_{0.8}$Y$_{0}.2$InCu$_{4}$. \cite{YbInCu4}
Magnetism in FeSb$_{2}$ in analogy to FeSi has been predicted by \textit{%
LDA+U} calculations. \cite{Lukoyanov} Besides the use of an external
magnetic field, one interesting possibility would be to induce the
ferromagnetic state by lattice expansion and band narrowing, as in FeSi$%
_{1-x}$Ge$_{x}$. \cite{Sunmog}$^{,}$ \cite{Anisimov3} Unfortunately,
isoelectronic lattice expansion is limited to rather small values of $x$ in
FeSb$_{2-x}$Bi$_{x}$. Our preliminary data show that the ferromagnetic state
is not reached for $x=0.016$. As in Fe$_{1-x}$Co$_{x}$Si, ferromagnetic
state is induced with Co substitution in FeSb$_{2}$. In both alloy systems
critical temperature T$_{C}$ exhibits a characteristic peak as a function of
Co concentration. \cite{Rongwei}$^{,}$\cite{Grigoriev} Whereas metallicity
simultaneously appears with ferromagnetism in Fe$_{1-x}$Co$_{x}$Si at $x=0.05
$, \cite{Dave} in Fe$_{1-x}$Co$_{x}$Sb$_{2}$ alloys transport and spin gap
vanish at $x=0.1$ and $x=0.2$ respectively. \cite{Rongwei}

What could be the mechanism of the WFM in Fe$_{1-x}$Co$_{x}$Sb$_{2}$ alloys?
Knowing that there is an inversion symmetry at the Fe site in the \textit{%
Pnnm} space group of FeSb$_{2}$, we can exclude the presence of the
Dzayloshinskii - Moriya (DM) type of interactions. This is in contrast to
the doped FeSi where the DM interaction is believed to be responsible for
the WFM. \cite{DM}$^{,}$\cite{Anisimov3} A canted antiferromagnetism can be
excluded based on the observed field dependence of the transition
temperature. That is, the ferromagnetic tail at low temperature is
insensitive to variation of the applied field. However, it is possible to
ascribe the low magnetic moment in Co doped FeSb$_{2}$ to the partial
ordering of Co$^{2+}$ ions. This scenario is in agreement with detailed
analysis of the magnetic and thermodynamic properties of Fe$_{1-x}$Co$_{x}$Sb%
$_{2}$. \cite{Rongwei}

Besides the obvious lattice expansion, the effect of the Co insertion is to
introduce extra carriers in the system. The carriers cause a closing of the
gap by $x=0.1$. \cite{Rongwei} Thus, the WFM appearance could be a
consequence of carrier-induced metallicity. This claim is further supported
by discarding another well known scenario for the WFM induction. More
precisely, one can imagine that the WFM is induced by an \textquotedblleft
inverted metal-insulator\textquotedblright\ scenario. \cite{Anisimov2}$^{,}$%
\cite{Anisimov3} In this scenario, magnetic order exists only in the
metallic phase. Furthermore, the metallicity is a direct consequence of
transition to the ferromagnetic state where a bulk moment of $\sim $ 1$\mu
_{B}$ develops out of small gap semiconductor with small susceptibility. 
\cite{Anisimov2}$^{,}$\cite{Anisimov3} However, in Fe$_{1-x}$Co$_{x}$Sb$_{2}$
for $x=0.2-0.45$ a small ordered moment is induced. Therefore, the presence
of the small moment excludes the \textquotedblleft inverted
metal-insulator\textquotedblright\ scenario and leaves as the only
possibility that the WFM arises as a consequence of carrier-induced
metallicity.

In conclusion, detailed structural and magnetic measurements argue against
extrinsic sources of WFM in Co - substituted FeSb$_{2}$. The ordered moment
below the WFM transition for Fe$_{0.75}$Co$_{0.25}$Sb$_{2}$ is $M\sim
(0.5-10)\cdot 10^{-3}\mu _{B}/Fe$. As opposed to FeSi where the metallic
state is caused by band narrowing of nearly ferromagnetic parent electronic
structure, weak ferromagnetism in Fe$_{1-x}$Co$_{x}$Sb$_{2}$ could be a
consequence of carrier induced metallic state. In order to fully understand
the magnetic structure, magnitude of moments, and mechanism of magnetic
ordering, further neutron scattering and/or nuclear magnetic resonance
measurements are envisaged.

We thank Yasutomo Uemura, Paul Canfield, T. M. Rice and Maxim Dzero for
useful communications and Dr. L. Rebbouh for assistance in the M\"{o}ssbauer
spectral measurements. This work was carried out at the Brookhaven National
Laboratory which is operated for the U.S. Department of Energy by Brookhaven
Science Associates (DE-Ac02-98CH10886), and at Department of Physics,
Universit\'{e} de Li\`{e}ge, Belgium. This work was also supported in part
by the National Science Foundation DMR-0547938 (V. F. M.).

$^{\ast }$Present address: Institut f\"{u}r Festk\"{o}rperforschung,
Forschungzentrum J\"{u}lich GmbH, D-52425 J\"{u}lich, Germany

\end{document}